\documentclass[twocolumn,floatfix,preprintnumbers,nofootinbib,superscriptaddress]{revtex4}

\usepackage{ulem}
\usepackage{bm}
\usepackage{times}
\usepackage{amssymb,amsbsy,amsmath,amsfonts}
\usepackage{graphicx}
\usepackage{float}
\usepackage{color}
\usepackage{morefloats}
\usepackage{rotating}
\usepackage{srcltx}
\usepackage{slashed}
\usepackage{subfigure}
\usepackage{multirow}
\usepackage{verbatim}
\usepackage{hyperref}
\usepackage{tabularx}
\usepackage{braket}

\usepackage[outdir=./]{epstopdf}

\usepackage{graphicx}
\usepackage{amsmath}
\usepackage{amsfonts}
\usepackage{amssymb}
\usepackage{color}
\usepackage{multirow}
\usepackage{mathrsfs}
\usepackage{gensymb}

\usepackage{booktabs}  
\usepackage{threeparttable}

\usepackage{subfigure}

\newcommand\be{\begin{eqnarray}}
\newcommand\ee{\end{eqnarray}}

\begin{document}

\title{The $\Sigma$ and $\Xi$ electromagnetic form factors in the extended vector meson dominance model}

\author{Bing Yan}~\email{yanbing@impcas.ac.cn}
\affiliation{College of Mathematics and Physics, Chengdu University of Technology, Chengdu 610059, China}
\affiliation{Institute of Modern Physics, Chinese Academy of Sciences, Lanzhou 730000, China}

\author{Cheng Chen}~\email{chencheng22@mails.ucas.ac.cn}
\affiliation{Institute of Modern Physics, Chinese Academy of Sciences, Lanzhou 730000, China}
\affiliation{School of Nuclear Sciences and Technology, University of Chinese Academy of Sciences, Beijing 101408, China}

\author{Ju-Jun Xie}~\email{xiejujun@impcas.ac.cn}
\affiliation{Institute of Modern Physics, Chinese Academy of Sciences, Lanzhou 730000, China}
\affiliation{School of Nuclear Sciences and Technology, University of Chinese Academy of Sciences, Beijing 101408, China}
\affiliation{Southern Center for Nuclear-Science Theory (SCNT), Institute of Modern Physics, Chinese Academy of Sciences, Huizhou 516000, Guangdong Province, China}

\begin{abstract}

We propose a phenomenological extended vector meson dominance model for the baryon electromagnetic structure, and it is found that the current experimental data on the $\Sigma$ and $\Xi$ electromagnetic form factors in the time-like region can be well described. Meanwhile, we can also reproduce the ratios of the total cross sections of reactions $e^+e^- \to \Sigma^+\bar{\Sigma}^-$, $\Sigma^0\bar{\Sigma}^0$, and $\Sigma^-\bar{\Sigma}^+$, which are $9.7 \pm 1.3 : 3.3 \pm 0.7 :1$ at center-of-mass energies from $2.3864$ to $3.02$ GeV. We also analytically continue the expression of the form factors to space-like region and estimate the charge radii of the $\Sigma$ and $\Xi$ hyperons. The result for the $\Sigma^-$ is in agreement with the experimental date.

\end{abstract}

\maketitle

\section{Introduction}

The electromagnetic structure information of hadrons is characterized by the electromagnetic form factors (EMFFs), which are functions of the four-momentum transfer squared $q^2$, with $q$ the four-momentum carried by the exchanged virtual photon. Study of these EMFFs can lead to a better understanding of fundamental structure of hadrons. On the experimental side, most commonly the baryon EMFFs in the space-like region ($q^2 < 0$) were measured in the electron-baryon scattering~\cite{SELEX:2001fbx,WA89:1999uls,JeffersonLabHallA:1999epl,JeffersonLabHallA:2001qqe}. While for these unstable hadrons, for example, these hyperons, their EMFFs in the space-like region are very difficult to be experimentally measured. However, in the time-like region ($q^2 > 0$), their EMFFs can be measured through the electron-positron annihilation reactions by the BESIII and Belle collaborations~\cite{BESIII:2017hyw,BESIII:2019cuv,BESIII:2020ktn,BESIII:2021aer,BESIII:2020uqk,BESIII:2021rkn,Wang:2022zyc,Belle:2022dvb}. Meanwhile, the effective form factor $G_{\rm eff}(q^2)$ of hyperons can be extracted from the high-precision measured Born cross sections of the reactions $e^+e^- \to Y\bar{Y}$ ($Y$ stands for hyperon; $\bar{Y}$ is anti-hyperon). It was pointed out that these baryon EMFFs in the time-like region can be associated with the time evolution of the charge and magnetic distributions inside the baryon~\cite{Belushkin:2006qa,Kuraev:2011vq}.

The hyperon effective form factors $G_{\rm eff}(q^2)$ are the functions that parametrize the $\gamma Y \bar{Y}$ vertex generated by the strong interaction. While, the production vertex $\gamma Y\bar{Y}$ is very poorly understood so far~\cite{Lorenz:2015pba,Mangoni:2021qmd}. The vector meson dominance (VMD) model is a very successful tool for studying the nucleon electromagnetic form factors, in both the space-like and time-like regions~\cite{Iachello:1972nu,Iachello:2004aq,Bijker:2004yu}. Within a modified VMD model, the EMFFs of $\Lambda$ hyperon were investigated in Refs.~\cite{Yang:2019mzq,Li:2021lvs}. By considering the $Y\bar{Y}$ final sate interactions, the EMFFs of hyperons in the time-like region have been studied in Ref.~\cite{Haidenbauer:2020wyp}. It is worth to mention that the enhancement of the effective form factor of the $\Lambda$ hyperon seen in the $e^+e^- \to \Lambda \bar{\Lambda}$ reaction, was reproduced within the two above different calculations in Refs.~\cite{Yang:2019mzq,Li:2021lvs} and Ref.~\cite{Haidenbauer:2020wyp}, respectively. In the vector meson dominance model for studying the electromagnetic form factors of baryons, there is a phenomenological intrinsic form factor $g(q^2)$. From these studies of the nucleon and hyperon EMFFs~\cite{Iachello:1972nu,Tomasi-Gustafsson:2001wyw,Iachello:2004aq,Bijker:2004yu,Haidenbauer:2020wyp,Bianconi:2015owa,Yang:2019mzq,Li:2021lvs,Bianconi:2015vva,BESIII:2021tbq,Dai:2021yqr,Yang:2022qoy,Cao:2021asd}, it is found that a better choice of $g(q^2)$ is the dipole form
\begin{eqnarray}
g(q^2) = \frac{1}{(1-\gamma q^2)^2},
\end{eqnarray}
with $\gamma$ a free parameter. In the space-like region, the dipole form is consistent with the results obtained from perturbative quantum chromodynamics calculations~\cite{Lepage:1979za,Lepage:1980fj}. In the time-like region, it should be noticed that $\gamma$ is a positive parameter, thus $g(q^2)$ will have a pole in the position $\gamma = 1/q^2$, such pole could be restricted in the unphysical region, if $\gamma$ satisfy $\gamma > 1/(4m^2_Y)$ for hyperon $Y$.

For a long time, the simple dipole form paremetrization is very useful for the discussion of different baryons. For example, the dipole form of $g(q^2)$ can well describe the effective form factors of $\Lambda$~\cite{Yang:2019mzq,Li:2021lvs}, $\Sigma$~\cite{Li:2020lsb}, and $\Xi$~\cite{Dai:2021yqr}. While for the nucleon, a good general review is given in Refs.~\cite{Perdrisat:2006hj,Denig:2012by,Pacetti:2014jai,Punjabi:2015bba}, both from the theoretical and from the experimental points of view. However, these determined values of $\gamma$ for different ground state octet baryons with spin $1/2$ are much different, even for the triplet $\Sigma^+$, $\Sigma^-$ and $\Sigma^0$~\cite{Li:2020lsb}. The determined values of $\gamma$, from previous works, for nucleon, $\Lambda$, $\Sigma$, and $\Xi^0$ baryons are collected in Table~\ref{tab:gammavalues}. Nevertheless, the VMD model and the parametrization of $g(q^2)$ can give a reasonable description of the experimental data on the baryon EMFFs at the considered energy region.

\begin{table}[htbp]
	\begin{center}
		\caption{\label{tab:azeroparameters} Values of $\gamma$ (in ${\rm GeV}^{-2}$) for octet baryons used in previous works.}
		\begin{tabular}{ccccc}
 \toprule[1.5pt]	
			 & Proton (\cite{Iachello:1972nu,Iachello:2004aq,Bijker:2004yu}) & Neutron (\cite{Bianconi:2015owa,Bianconi:2015vva,BESIII:2021tbq,Dai:2021yqr})  & $\Lambda$ (\cite{Yang:2019mzq}) & $\Lambda$ (\cite{Li:2021lvs})   \\ \hline
		$\gamma$  & $1.408$     & $1.408$   & $0.336$   & $0.48 \pm 0.08$ \\ \hline 
               &$\Sigma^+$ (\cite{Li:2020lsb}) & $\Sigma^-$ (\cite{Li:2020lsb}) & $\Sigma^0$ (\cite{Dai:2021yqr}) & $\Xi^0$ (\cite{Dai:2021yqr})  \\ \hline
		$\gamma$  & $0.46 \pm 0.01$ & $1.18 \pm 0.13$   &  $0.26 \pm 0.01$   & $0.21 \pm 0.02$     \\
       \toprule[1.5pt]
		\end{tabular} \label{tab:gammavalues}
	\end{center}
\end{table}

Various experimental and theoretical efforts have been contributed to the electromagnetic form factors. Very recently, the EMFFs of $\Sigma^+$, $\Sigma^-$, and $\Sigma^0$ hyperons in the time-like region, have been measured with high-precision by the BESIII collaboration through $e^+e^- \to \Sigma^+\bar{\Sigma}^-$~\cite{BESIII:2020uqk}, $\Sigma^-\bar{\Sigma}^+$~\cite{BESIII:2020uqk}, and $\Sigma^0\bar{\Sigma}^0$ reactions~\cite{BESIII:2021rkn} at center-of-mass energies from $2.3864$ to $3.02$ GeV. The resulting ratios of total cross sections of these above three reactions are $9.7 \pm 1.3$ : $1$ : $3.3 \pm 0.7$~\cite{BESIII:2020uqk,Mangoni:2021lmr,Irshad:2022zga}, which disagree with various theoretical model predictions~\cite{Anselmino:1992vg,Kubis:2000aa}. After the experimental measurements of $e^+e^- \to \Sigma^+\bar{\Sigma}^-$ and $\Sigma^-\bar{\Sigma}^+$~\cite{BESIII:2020uqk}, the effective form factors of $\Sigma^+$ and $\Sigma^-$ were investigated by using the VMD model~~\cite{Li:2020lsb}, where the parameter $\gamma$ were taken with different values for $\Sigma^+$ and $\Sigma^-$. In Ref.~\cite{Haidenbauer:2020wyp}, by considering the final state interactions of $Y\bar{Y}$, the energy dependence of the three reactions $e^+e^- \to \Sigma^+\bar{\Sigma}^-$, $\Sigma^0\bar{\Sigma}^0$, and $\Sigma^-\bar{\Sigma}^+$ at low energies can be roughly reproduced, and it was found that there is a strong interplay between $\Sigma^+\bar{\Sigma}^-$, $\Sigma^0\bar{\Sigma}^0$, and $\Sigma^-\bar{\Sigma}^+$ channel in the near-threshold region, caused by the $\Sigma\bar{\Sigma}$ final state interactions.

In the present work, we revisit the EMFFs at the time-like region of $\Sigma$ and $\Xi$ hyperons within an extended vector meson dominance model, where the affects of the isospin combinations from isovector $\rho^0$ and isoscalar $\omega$ and $\phi$ mesons are taken into account. Furthermore, we assume that the values of model parameter $\gamma$ are same for $\Sigma$ and $\Xi$ hyperons. In addition, a vector meson with mass around 2.7 GeV was considered for the sake of better fitting the EMFFs of the $\Xi^0$ and $\Xi^-$ hyperons. We then progress to an analysis of the electromagnetic form factors in the space-like region and evaluate the electromagnetic radius of $\Sigma$ hyperons. The theoretical result for the $\Sigma^-$ hyperon is in agreement with the experimental measurements.

This article is organized as follows: in next section we will show the theoretical formalism of the $\Sigma$ and $\Xi$ electromagnetic form factors in the VMD model. Numerical results about the effective form factors of $\Sigma$ and $\Xi$ and total cross sections of $e^+e^- \to \Sigma\bar{\Sigma}$ and $\Xi\bar{\Xi}$ are shown in Sec.~\ref{sec:results}, and a short summary is given in the final section.

\section{Formalism} \label{sec:formalism}

As already pointed out, as fixed-energy $e^+e^-$ colliders, the EMFFs of hyperons in the time-like region was extracted from the data on the differential cross section of the process $e^+e^- \to Y\bar{Y}$. For analysis the data, the BESIII Collaboration use the energy scan method~\cite{BESIII:2019hdp,Xia:2021agf,Zhou:2022jwr}, while the initial state radiation method was used by Belle Collaboration~\cite{Belle:2022dvb} and $BABAR$ collaboration~\cite{BaBar:2005pon,BaBar:2007fsu}. Besides, the effective form factors $G_{\rm eff}$ can be easily obtained from the data of the total cross sections. The module squared of effective form factor $|G_{\rm eff}|^2$ is a linear combination of $|G_E|^2$ and $|G_M|^2$, and proportional to the total cross section of $e^+e^- \to Y\bar{Y}$ reaction.
In this work, we study the EMFFs of $\Sigma$ and $\Xi$ baryons in the time-like region with the experimental measurements on the $e^+ e^- \to Y\bar{Y}$ reactions. Based on parity conservation and Lorentz invariant, the electromagnetic current of the baryons with a spin of $1/2$ characterize two independent scalar functions $F_1(q^2)$ and $F_2(q^2)$ depending on $q^2$, which are the Dirac and Pauli form factors, respectively. While the corresponding electrical and magnetic form factors $G_E(q^2)$ and $G_M(q^2)$ are written as~\cite{Sachs:1962zzc,Green:2014xba,Irshad:2022zga},
\begin{eqnarray}
    G_E(q^2) &=& F_1(q^2) + \tau F_2(q^2), \\
    G_M(q^2) &=& F_1(q^2) + F_2(q^2),
\end{eqnarray}
where $M$ is the baryon mass and $\tau = q^2/(4M^2)$. With $G_E(q^2)$ and $G_M(q^2)$, the magnitude of the effective form factor $|G_{\rm eff}(q^2)|$ is defined as
\begin{eqnarray}
| G_{\rm eff} (q^2) | = \sqrt{\frac{2\tau|G_M(q^2)|^2 + | G_E(q^2)|^2}{1+2\tau}}.
\end{eqnarray}

\begin{figure}[htbp]
    \centering
    \includegraphics[scale=0.38]{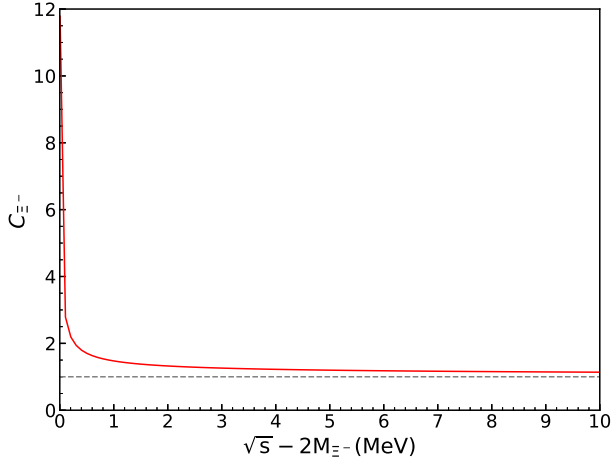}
    \caption{The Coulomb factor for $\Xi^-$. The dashed horizontal line stands for $C_{\Xi^-} = 1$.}
    \label{fig:CY}
\end{figure}

In the time-like region, the effective form factors of hyperons are experimentally studied via the electron-positron annihilation processes. Under the one photon exchange approximation, the total cross section of the annihilation reaction $e^{+}e^{-} \to \bar{Y} Y$ can be expressed in terms of the effective form factor $G_{\rm eff}$ as~\cite{Dobbs:2014ifa,BaBar:2005pon,Haidenbauer:2014kja}
\begin{equation}
\sigma_{e^{+}e^{-}\rightarrow\bar{Y}Y} =\frac{4\pi\alpha^{2}\beta C_Y}{3s}(1+\frac{1}{2\tau})\mid G_{\rm eff} (s)\mid^2 ,
\end{equation}
with $\alpha = e^2/(4\pi)=1/137.036$ the fine-structure constant, and $\beta=\sqrt{1-4M_Y^2/s}$ is a phase-space factor. Here, $s=q^2$ is the invariant mass square of the $e^+e^-$ system. The coulomb enhancement factor $C_Y$~\footnote{It is also called Sommerfeld factor.} accounts for the electromagnetic interaction of charged point-like fermion pairs in the final state~\cite{Arbuzov:2011ff}, which is given by
\begin{eqnarray}
C_Y=\begin{cases}
\frac{y}{1-e^{-y}}  & ~~{\rm for} ~\Sigma^+,~\Sigma^-,~{\rm and}~ \Xi^-,\\
   1   & ~~{\rm for}~\Sigma^0 ~{\rm and}~ \Xi^0,
\end{cases}
\end{eqnarray}
with $y = \frac{\alpha \pi}{\beta} \frac{2M_Y}{\sqrt{s}}$. Considering the $C_Y$ factor, it is expected that the cross section of process $e^+e^- \to Y\bar{Y}$ is nonzero at the reaction threshold for charged hyperons pairs. As plotted in Fig.~\ref{fig:CY} for the case of $\Xi^-$~\footnote{The numerical results for $\Sigma^+$ and $\Sigma^-$ are similar.}, one can see that the factor $C_Y$ affects only at the energy region very close to the reaction threshold. Moreover, it decreases very quickly as the reaction energy growing and it follows that few MeV above the reaction threshold it is $C_Y \sim 1$, then its effect can be safely neglected~\cite{Baldini:2007qg,BaldiniFerroli:2010ruh,Arbuzov:2011ff,Cao:2019wwt}.

\subsection{The EMFFs of $\Sigma$ hyperon}

In the VMD model, for the $e^+e^- \to \Sigma\bar{\Sigma}$ reaction, the virtual photon couples to $\Sigma$ and $\bar{\Sigma}$ through isovector $\rho^0$ meson and isoscalar $\omega$ and $\phi$ mesons. Since both the $\omega$ and $\phi$ are far from the mass threshold of $\Sigma\bar{\Sigma}$, the behavior of the contributions from them are similar, thus we combine their contributions. In this way, one can parameterize Dirac and Pauli form factors for $\Sigma^+$ and $\Sigma^-$ in the time-like region as follows~\cite{Iachello:1972nu,Bijker:2004yu},~\footnote{We have followed:
\begin{eqnarray}
\ket{\Sigma^+\bar{\Sigma}^-}&=&\frac{1}{\sqrt{2}}\ket{1,0}+\frac{1}{\sqrt{3}}\ket{0,0}+\frac{1}{\sqrt{6}}\ket{2,0}, \nonumber \\
\ket{\Sigma^-\bar{\Sigma}^+}&=&-\frac{1}{\sqrt{2}}\ket{1,0}+\frac{1}{\sqrt{3}}\ket{0,0}+\frac{1}{\sqrt{6}}\ket{2,0}, \nonumber \\
\ket{\Sigma^0\bar{\Sigma}^0}&=&-\frac{1}{\sqrt{3}}\ket{0,0}+\sqrt{\frac{2}{3}}\ket{2,0} ,\nonumber\\
\end{eqnarray}
with the basis of $\ket{I_{\Sigma\bar{\Sigma}},I^Z_{\Sigma\bar{\Sigma}}}$. In the one photon exchange approximation, there is no contributions from the isospin tensor terms.}
\begin{eqnarray}
    F_1^{\Sigma^{+}}&=&g(q^{2})(f_1^{\Sigma^{+}}+\frac{\beta_\rho}{\sqrt{2}} B_\rho - \frac{\beta_{\omega\phi}}{\sqrt{3}} B_{\omega\phi}) , \\
    F_2^{\Sigma^{+}} &=& g(q^{2})(f_2^{\Sigma^{+}} B_\rho -  \frac{\alpha_{\omega\phi}}{\sqrt{3}}B_{\omega\phi}), \\
    F_1^{\Sigma^{-}} &=& g(q^{2})(f_1^{\Sigma^{-}}-\frac{\beta_\rho}{\sqrt{2}}B_\rho -  \frac{\beta_{\omega\phi}}{\sqrt{3}}B_{\omega\phi}), \\
    F_2^{\Sigma^{-}} &=& g(q^{2})(f_2^{\Sigma^{-}}B_\rho-  \frac{\alpha_{\omega\phi}}{\sqrt{3}}B_{\omega\phi}), \\
    F_1^{\Sigma^{0}} &=& g(q^{2})(\frac{\beta_{\omega\phi}}{\sqrt{3}}-\frac{\beta_{\omega\phi}}{\sqrt{3}}B_{\omega\phi}), \\
    F_2^{\Sigma^{0}} &=& g(q^{2}) \mu_{\Sigma^{0}} B_{\omega \phi},
\end{eqnarray}
with
\begin{eqnarray}
B_\rho  &=& \frac{m_{\rho}^{2}}{m_{\rho}^{2}-q^{2} - i m_{\rho}\Gamma_{\rho}} , \\
B_{\omega \phi} &=& \frac{m_{\omega\phi}^{2}}{m_{\omega\phi}^{2}-q^{2}- im_{\omega\phi}\Gamma_{\omega\phi}},
\end{eqnarray}
where the widths of $\rho$, $\omega$ and $\phi$ are taken into account. In this work, we take $m_{\rho}=0.775$ MeV, $\Gamma_{\rho}=149.1\ \rm MeV,\ \Gamma_{\omega\phi}=(\Gamma_{\omega}+\Gamma_{\phi})/2=6.4645\ \rm MeV$, and $\ m_{\omega\phi}=(m_{\omega}+m_{\phi})/2=0.9005\ \rm GeV$, which are quoted in the review of particle physics book~\cite{ParticleDataGroup:2022pth}. Besides, we take $\mu_{\Sigma^{+}}=3.112\hat{\mu}_{\Sigma^{+}}$, $\mu_{\Sigma^{-}}=-1.479\hat{\mu}_{\Sigma^{-}}$, $\mu_{\Sigma^{0}}=2.044\hat{\mu}_{\Sigma^{0}}$ in natural unit~\cite{ParticleDataGroup:2022pth}, i.e., $\hat{\mu}=\frac{e}{2M_{\Sigma}}$. In addition, at $q^2=0$, with the constraints $G_E^{\Sigma^{+}}=1$ and $G_M^{\Sigma^{+}}=\mu_{\Sigma^{+}}$, $G_E^{\Sigma^{-}}=-1$ and $G_M^{\Sigma^{-}}=\mu_{\Sigma^{-}}$, the coefficients $f_1^{\Sigma^{+}}$ and $f_2^{\Sigma^{+}}$, $f_1^{\Sigma^{-}}$ and $f_2^{\Sigma^{-}}$ can be calculated,
\begin{eqnarray}
f_1^{\Sigma^+} \!\!\! &=& 1-\frac{\beta_\rho}{\sqrt{2}}+\frac{\beta_{\omega \phi} }{\sqrt{3}},~~~ f_2^{\Sigma^+} = 2.112+\frac{\alpha_{\omega \phi}}{\sqrt{3}}, \\
f_1^{\Sigma^-} \!\! \! &=& -1+\frac{\beta_\rho}{\sqrt{2}}+\frac{\beta_{\omega \phi} }{\sqrt{3}},~~~f_2^{\Sigma^-} = -0.479+\frac{\alpha_{\omega \phi}}{\sqrt{3}}. 
\end{eqnarray}

Finally, the model parameters $\gamma$, the coefficients $\beta_\rho$, $\beta_{\omega\phi}$, and $\alpha_{\omega\phi}$ will be determined by fitting them to the experimental data on the time-like effective form factors of $\Sigma^+$, $\Sigma^0$, and $\Sigma^-$, which will be discussed in following.

\subsection{The EMFFs of $\Xi$ hyperon}

For the case of $e^+e^- \to \Xi^-\bar{\Xi}^+$ and $\Xi^0\bar{\Xi}^0$ reactions, since $\Xi^-$ and $\Xi^0$ are isospin doublets, we express the $\Xi^-\bar{\Xi}^+$ and $\Xi^0\bar{\Xi}^0$ states in terms of isospin 0 and 1 components. The mixtures of isoscalar and isovector for $\Xi^-\bar{\Xi}^+$ and $\Xi^0\bar{\Xi}^0$ of equal relative wight but different sign are imposed by the isospin symmetry as introduced by the underlying Clebsch-Gorden coefficients~\cite{ParticleDataGroup:2022pth}. Then, the Dirac and Pauli form factors $F_1$ and $F_2$ for $\Xi^-$ and $\Xi^0$ can be easily obtained as before for the $\Sigma$ hyperon,
\begin{eqnarray}
    F_1^{\Xi^-} &=& g(q^2)(f_1^{\Xi^-} - \frac{\beta_{\rho}}{\sqrt{2}} B_\rho- \frac{\beta_{V_1}}{\sqrt{2}}B_{V_1}  \nonumber \\ 
    &&  
     -\frac{\beta_{V_2}}{\sqrt{2}}B_{V_2}+ \frac{\beta_{\omega\phi}}{\sqrt{2}}B_{\omega\phi} ), \\
    F_2^{\Xi^-} &=& g(q^2)(f_2^{\Xi^-}B_\rho-\frac{\alpha_{V_1}}{\sqrt{2}}B_{V_1} \nonumber \\
    && -\frac{\alpha_{V_2}}{\sqrt{2}}B_{V_2}+\frac{\alpha_{\omega \phi}}{\sqrt{2}}B_{\omega \phi}), \\
    F_1^{\Xi^0} &=& g(q^2)(f_1^{\Xi^0}+\frac{\beta_\rho}{\sqrt{2}} B_\rho+\frac{\beta_{V_1}}{\sqrt{2}}B_{V_1} \nonumber \\
    &&  +\frac{\beta_{V_2}}{\sqrt{2}}B_{V_2}+\frac{\beta_{\omega \phi}}{\sqrt{2}}B_{\omega \phi}), \\
    F_2^{\Xi^0} &=& g(q^2)(f_2^{\Xi^0}B_\rho+\frac{\alpha_{V_1}}{\sqrt{2}}B_{V_1} \nonumber \\
    && +\frac{\alpha_{V_2}}{\sqrt{2}}B_{V_2}+\frac{\alpha_{\omega \phi}}{\sqrt{2}}B_{\omega \phi}),
\end{eqnarray}
with
\begin{eqnarray}
B_{V1} &=& \frac{M_{V_1}^2}{M_{V_1}^2-q^2-i M_{V_1}\Gamma_{V_1}},\\
B_{V2} &=& \frac{M_{V_2}^2}{M_{V_2}^2-q^2-i M_{V_2}\Gamma_{V_2}},
\end{eqnarray}
where we have considered contributions from two more excited vector mesons, $V_1$ and $V_2$, in addition the contributions from ground states $\rho$, $\omega$ and $\phi$. Their mass and width are $M_{V_1}$ ($M_{V_2}$) and $\Gamma_{V_1}$ ($\Gamma_{V_2}$), respectively. The mass $M_{V_2}$ and width $\Gamma_{V_2}$ are taken as used in Ref.~\cite{BESIII:2020ktn}, which are: $M_{V_2}=2.993\ \rm GeV$ and $\Gamma_{V_2}= 88\ {\rm MeV}$. Besides, we take ${\mu}_{\Xi^-}=-0.915\hat{\mu}_{\Xi^{-}}$, and ${\mu}_{\Xi^0}=-1.749\hat{\mu}_{\Xi^{0}}$ in natural unit~\cite{ParticleDataGroup:2022pth}. Then the coefficients $f_1^{\Xi^-}$, $f_2^{\Xi^-}$, $f_1^{\Xi^0}$, and $f_2^{\Xi^0}$ can be calculated as
\begin{eqnarray}
 f_1^{\Xi^-} &=& -1+\frac{\beta_{\rho}}{\sqrt{2}}+\frac{\beta_{V_1}}{\sqrt{2}}+\frac{\beta_{V_2}}{\sqrt{2}}-\frac{\beta_{\omega\phi}}{\sqrt{2}},\\
 f_2^{\Xi^-} &=& 0.085+\frac{\alpha_{V_1}}{\sqrt{2}}+\frac{\alpha_{V_2}}{\sqrt{2}}-\frac{\alpha_{\omega\phi}}{\sqrt{2}}, \\
 f_1^{\Xi^0} &=& -\frac{\beta_{\rho}}{\sqrt{2}}-\frac{\beta_{V_1}}{\sqrt{2}}-\frac{\beta_{V_2}}{\sqrt{2}}-\frac{\beta_{\omega\phi}}{\sqrt{2}}, \\
 f_2^{\Xi^0} &=&  -1.749-\frac{\alpha_{V_1}}{\sqrt{2}}-\frac{\alpha_{V_2}}{\sqrt{2}}-\frac{\alpha_{\omega\phi}}{\sqrt{2}}.
\end{eqnarray}

The parameter $\gamma$ will be fixed as the one determined from the case of $\Sigma$, while the other free parameters $\beta_{\omega \phi}$, $\beta_{\rho}$, $\beta_{V_1}$, $\beta_{V_2}$, $\alpha_{\omega \phi}$, $\alpha_{V_1}$, $\alpha_{V_2}$, $\Gamma_{V_1}$, and $M_{V_1}$ are determined by fitting them to experimental data on the time-like effective form factors of $\Xi^-$ and $\Xi^0$.

\section{Numerical results} \label{sec:results}

\begin{figure}[htbp]
    \centering
    \includegraphics[scale=0.38]{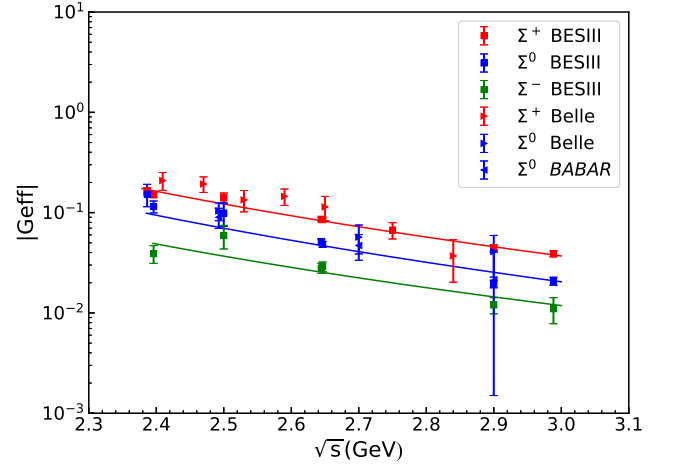}
    \caption{The obtained effective form factors of $\Sigma^+$, $\Sigma^0$, and $\Sigma^-$, compared with the experimental data.}
    \label{fig:GeffSigma}
\end{figure}

\begin{figure}[htbp]
    \centering
    \includegraphics[scale=0.38]{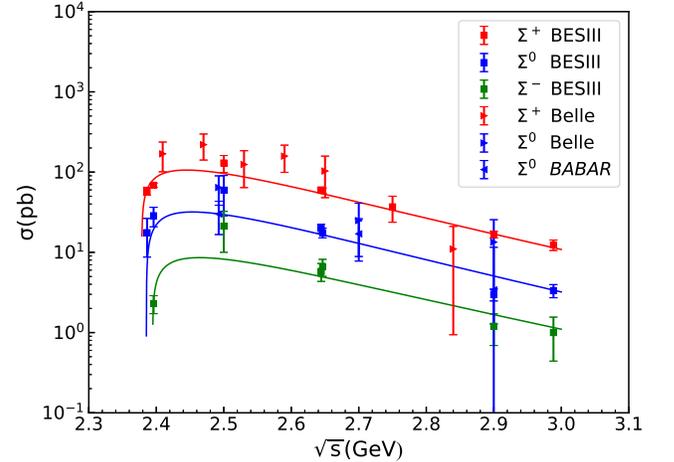}
    \caption{The total cross section of $\Sigma^+$, $\Sigma^0$ and $\Sigma^-$ hyperons compared with experimental data. }
    \label{fig:tcsSigma}
\end{figure}

Under the above formulations, we perform a four-parameter $(\gamma,\ \beta_\rho,\ \beta_{\omega\phi},\ \alpha_{\omega\phi})$-$\chi^2$ fit to the experimental data on the effective form factors $G_{\rm eff}$ of $\Sigma^+$, $\Sigma^0$, and $\Sigma^-$ hyperons. There are $33$ data points in total, which are extracted at the center-of-mass energies from $2.3864$ to $3.0200\ \rm GeV$. The fitted parameters are: $\gamma  = 0.527 \pm 0.024$ ${\rm GeV}^{-2}$, $\beta_\rho = 1.63 \pm 0.07$, $\beta_{\omega\phi} = -0.08 \pm 0.06$, and $\alpha_{\omega\phi} = -3.18 \pm 0.77$. And the obtained $\chi^2/{\rm dof}$ is $1.69$, where ${\rm dof}$ is the number of dimension of the freedom. Note that the obtained $\chi^2 / {\rm dof}$ is larger than 1, since we have fitted all the experimental data from BESIII~\cite{BESIII:2020uqk,BESIII:2021rkn}, Belle~\cite{Belle:2022dvb}, and $BABAR$~\cite{BaBar:2007fsu} Collaborations, by considering these contributions from only ground state of vector mesons. If we considered only these data of BESIII Collaboration~\cite{BESIII:2020uqk,BESIII:2021rkn}, the obtained $\chi^2 / {\rm dof}$ is $1.17$. In Fig.~\ref{fig:GeffSigma} we show the theoretical results of the effective form factors of the $\Sigma^+$, $\Sigma^0$, and $\Sigma^-$. The red, blue, and green curves stand for the results for $\Sigma^+$, $\Sigma^0$, and $\Sigma^-$, respectively. The experimental data from BESIII~\cite{BESIII:2020uqk,BESIII:2021rkn}, Belle~\cite{Belle:2022dvb}, and $BABAR$ Collaboration~\cite{BaBar:2007fsu} are also shown for comparing. One can see that, with same model parameters, we can describe these data on the effective form factors of $\Sigma^+$, $\Sigma^0$ and $\Sigma^-$ quite well, especially for the precise data measured by the BESIII Collaboration~\cite{BESIII:2020uqk,BESIII:2021rkn}.
The total cross sections of $e^+ e^- \to \Sigma \bar{\Sigma}$ are also calculated with these fitted parameters. The numerical results are shown in Fig.~\ref{fig:tcsSigma}, compared with the experimental data. Since the effective form factors of $\Sigma$ hyperons can be well reproduced with our model, the total cross sections of $e^+e^- \to \Sigma^+ \bar{\Sigma}^-$, $e^+e^- \to \Sigma^0 \bar{\Sigma}^0$ and $e^+e^- \to \Sigma^-\bar{\Sigma}^+$ reactions can be also well described.

\begin{figure}[htbp]
   \centering
    \includegraphics[scale=0.38]{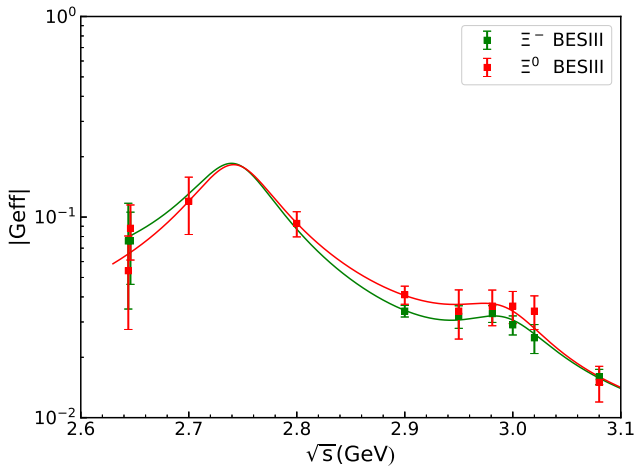}
    \caption{The obtained effective form factors of $\Xi^-$ and $\Xi^0$ compared with the experimental data.}
    \label{fig:GeffXi}
\end{figure}

\begin{figure}[htbp]
    \centering
    \includegraphics[scale=0.38]{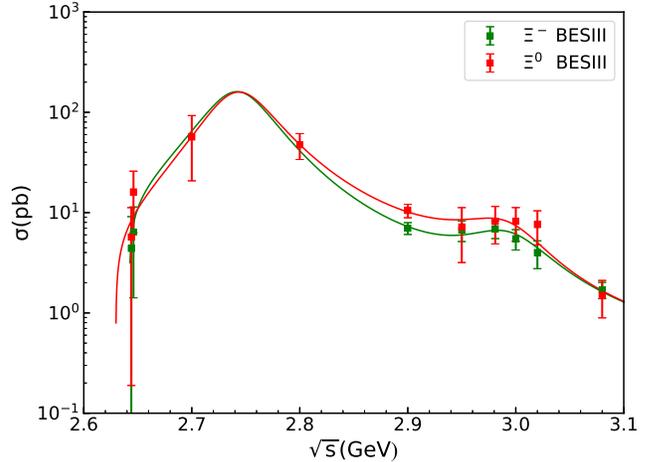}
    \caption{The total cross sections of $e^+e^- \to \Xi^- \bar{\Xi}^+$ and $e^+e^- \to \Xi^0 \bar{\Xi}^0$ reactions compared with experimental data. }
    \label{fig:tcsXi}
\end{figure}

\begin{table}[htbp]
    \centering
    \caption{Fitted model parameters for the effective form factors of $\Xi^-$ and $\Xi^0$.}
    \begin{tabular}{c | c | c | c}
    \toprule[1.5pt]
       Parameter  & Value & Parameter & Value \\
       \hline
        $\beta_{\omega \phi}$ & $-0.774$ & $\alpha_{\omega \phi}$ & $9.346$ \\
        $\beta_{\rho}$ & $0.616$ & $\alpha_{V_1}$ & $-0.039$ \\
        $\beta_{V_1}$ & $0.099$ & $\alpha_{V_2}$ & $-0.113$ \\
        $\beta_{V_2}$ & $0.115$ & $\Gamma_{V_1}$ (MeV) & $71$ \\
        $M_{V_1}$ (GeV) & $2.742$  & \\
        \toprule[1.5pt]
    \end{tabular}    \label{tab:fittedparametersXi}
\end{table}
For the case of $\Xi^-$ and $\Xi^0$ effective form factors, $\gamma$ is taken as the result of fitting to $\Sigma$ hyperon, i.e., $\gamma=0.527$, we perform nine-parameter $(\beta_{\omega \phi}$, $\beta_{\rho}$, $\beta_{V_1}$, $\beta_{V_2}$, $\alpha_{\omega \phi}$, $\alpha_{V_1}$, $\alpha_{V_2}$, $\Gamma_{V_1}$, $M_{V_1})$-$\chi^2$ fit to the experimental data on. There are totally 18 data points, and these data correspond to the center-of-mass energies from 2.644 to 3.080 GeV. The fitted parameters are listed in Table~\ref{tab:fittedparametersXi}, with a reasonably small $\chi^2/{\rm dof} = 0.29$. Since we have more free parameters and the experimental data points is limited, we did not get the uncertainties of these parameters from the $\chi^2$ fit. In Fig.~\ref{fig:GeffXi}, we depict the effective form factor of the $\Xi^-$ and $\Xi^0$ using the fitted parameters shown in Table~\ref{tab:fittedparametersXi}. The red curve stands for the results of $\Xi^0$, while the green curve is the fitted results for $\Xi^-$. Again, one can see that the experimental data on the effective form factors of $\Xi^-$ and $\Xi^0$ can be well reproduced. It is worth to mention that the two resonances $V_1$ and $V_2$ are crucial to describe the experimental data, and without their contributions, we cannot get a good fit to the experimental data. In addition, the total cross section of $e^+e^- \to \Xi^- \bar{\Xi}^+$ and $e^+ e^- \to \Xi^0\bar{\Xi}^0$ are also calculated with the fitted parameters shown in Table~\ref{tab:fittedparametersXi}, and the numerical results are shown Fig.~\ref{fig:tcsXi}. The two peaks of $V_1$ and $V_2$ can be clear seen, and more precise data around 2744 and 2993 MeV are needed to further study their properties.

We next pay attention to the EMFFs at the space-like region, which can be straightforwardly obtained with the these parameters determined from the experimental data in the time-like region. Since the EMFFs in the space-like region are real, thus we have to ignore the widths of the vector mesons. Then one can calculate the mean squared charge radius, which is defined by the relation~\cite{SELEX:2001fbx,Atac:2021wqj,Kubis:2000aa}
\begin{eqnarray}
{\left\langle r_{ch}^2  \right\rangle} =  \begin{cases}
    \frac{-6}{G_E(0)}{\frac{dG_E(Q^2)}{dQ^2}}\bigg|_{Q^2=0}, & ~ {\rm for}~ \Sigma^+, ~\Sigma^-~ {\rm and} ~\Xi^-,\\
     -6 {\frac{dG_E(Q^2)}{dQ^2}}\bigg|_{Q^2=0}, &  ~{\rm for} ~\Sigma^0 ~{\rm and}~ \Xi^0, 
\end{cases}
\end{eqnarray}
with $Q^2 = - q^2$. With the parameters fitted above, the calculated results of $ {\left\langle r_{ch}^2 \right\rangle}$ of $\Sigma$ 
and $\Xi$ hyperons are shown in Table~\ref{tab:radius}. Our result for $\Sigma^-$ is agreement with the experimental data within uncertainties:
 ${\left\langle r_{ch}^2 \right\rangle}_{\Sigma^-}=0.61 \pm 0.12 \pm 0.09$~\cite{SELEX:2001fbx}, ${\left\langle r_{ch}^2 \right\rangle}_{\Sigma^-}=0.91 \pm 0.32\pm 0.4$~\cite{WA89:1999uls}.
  In Ref.~\cite{SELEX:2001fbx} the $\Sigma^-$ charge radius was measured in the space-like $Q^2$ range $0.035-0.105~\rm{GeV}^2$ by elastic scattering of a $\Sigma^-$ beam off atomic electrons.
   The measurement was performed with the SELEX (E781) spectrometer using the Fermilab hyperon beam at a mean energy of $610 {\rm GeV}$. In Ref.~\cite{WA89:1999uls} it was attracted from the elastic 
   scattering of high energy $\Sigma^-$ off electrons from carbon and copper targets using the CERN hyperon beam, where these events are identified using a maximum likelihood technique exploring the 
   kinematical relations of the scattered particles. Theoretical calculations with chiral perturbation theory (ChPT)~\cite{Kubis:2000aa,HillerBlin:2017syu} and the nonlocal chiral effective theory (ChET)~\cite{Yang:2020rpi}, 
   and chiral constituent quark model (ChCQM)~\cite{Wagner:1998fi} are also listed for comparison. On can see that the orderings of the most
charge radii calculated by other works are in agreement with our results. Moreover, our results are consistent with these calculations in Refs.~\cite{HillerBlin:2017syu,Yang:2020rpi,Wagner:1998fi} that ${\left\langle r_{ch}^2 \right\rangle}_{\Sigma^+}>{\left\langle r_{ch}^2 \right\rangle}_{\Sigma^-}$. On the contrary, the results obtained with chiral perturbation theory predictions in Ref.~\cite{Kubis:2000aa} indicate that the charge radius of $\Sigma^- $ is larger than the one of $\Sigma^+ $. In addition, the charge radius of $\Xi^0$ calculated here is small and negative, which is in agreement with the nonlocal chiral effective theory calculation in Ref.~\cite{Yang:2020rpi}. It is expected that these results can be tested by future experimental measurements.
\begin{table*}[htbp]
\centering
 \caption{The obtained results for mean squared electromagnetic radii ${\left\langle r_{ch}^2 \right\rangle}$ ($\rm{fm}^2$) for $\Sigma$ and $\Xi$. The results from two ChPT calculations, ChET and, $\rm{ChCQM}$ as well as the experimental data are also listed.}
    \begin{tabular}{c | c | c | c | c | c }
    \toprule[1.5pt]
       Baryon  & $\Xi^0$ & $\Xi^-$ & $\Sigma^+$ & $\Sigma^0$ & $\Sigma^-$  \\ \hline
        This work &   $-0.07$ & $0.43$ & $0.78$  & $0.12$ & $0.65$   \\
        ChPT~\cite{Kubis:2000aa} & $0.13 \pm 0.03$  &  $0.49 \pm 0.05$ &  $0.60 \pm 0.02$ & $-0.03$ $\pm$ $0.01$ & $0.67 \pm 0.03$  \\
       ChPT~\cite{HillerBlin:2017syu} &   $0.36 \pm 0.02$  &  $0.61 \pm 0.01$ &  $0.99 \pm 0.03$ & $0.10$ $\pm$ $0.02$ & $0.780$   \\
       ChET~\cite{Yang:2020rpi} &    $-0.015 \pm 0.007$ & $0.601\pm 0.127$ & $0.719 \pm 0.116$ & 0.010 $\pm$ 0.004 & $0.700 \pm 0.124$  \\
       ChCQM~\cite{Wagner:1998fi}   &  $0.091$ & $0.587$ & $0.825$ & $0.089$ & $0.643$  \\    
       \toprule[1.5pt]
    \end{tabular}
    \label{tab:radius}
\end{table*}

\section{Summary}

In this work, we study the effective form factor of $\Sigma$ and $\Xi$ hyperons in time-like region within the vector meson dominance model, and we take a common model parameter $\gamma$. In addition, the effect of the isospin combination is taken into account. For the case of $\Sigma$ hyperon, the contributions from $\rho$, $\omega$ and $\phi$ mesons are considered. Within same model parameters, we can simultaneously describe the current experimental data on the effective form factors of $\Sigma^+$, $\Sigma^0$ and $\Sigma^-$. While for the case of $\Xi^+$ and $\Xi^-$, in addition to the contributions of the ground states $\rho$, $\omega$ and $\phi$, it is found that one needs also contributions from two new vector states, and their masses and widths are: $M_{V_1} = 2.742$ GeV, $\Gamma_{V_1} = 71$ MeV, $M_{V_2} = 2.993$ GeV, and $\Gamma_{V_2} =  88$ MeV. It is expected that new precise experimental data at BESIII~\cite{BESIII:2020nme} can be used to further study their properties.
Finally, we would like to stress that thanks to the effects of the isospin combinations, the effective form factors of $\Sigma^+$, $\Sigma^0$ and $\Sigma^-$ can be simultaneously reproduced within the same model parameters by using the vector meson dominance model. Again, the theoretical results obtained here also indicate that the vecor meson dominance model is a valid tool for studying the baryonic electromagnetic form factors at the time-like region. More precise data on the $e^+e^- \to Y\bar{Y}$ reactions can be used to improve our knowledge of hyperon effective form factors.

\section*{Acknowledgements}

We warmly thank Profs. Xiong-Fei Wang and Xiao-Rong Zhou for useful comments and discussions. This work is partly supported by the National Natural Science Foundation of China under Grant Nos. 12075288, 11735003, and 11961141012. It is also supported by the Youth Innovation Promotion Association CAS.

\end{document}